# Uncertainty Quantification of Mode Shape Variation Utilizing Multi-Level Multi-Response Gaussian Process

K. Zhou

Postdoctoral Researcher

J. Tang[†]

Professor

Department of Mechanical Engineering

University of Connecticut

191 Auditorium Road, Unit 3139

Storrs, CT 06269, USA

Phone: (860) 486-5911, Email: jiong.tang@uconn.edu

---

[†] Corresponding author

# Uncertainty Quantification of Mode Shape Information Utilizing Multi-level Multi-response Gaussian Process

K. Zhou and J. Tang[†],

Department of Mechanical Engineering

University of Connecticut

Storrs, CT 06269

USA

Phone: +1 (860) 486-5911, Email: jiong.tang@uconn.edu

**ABSTRACT**

Mode shape information play the essential role in deciding the spatial pattern of vibratory response of a structure. The uncertainty quantification of mode shape, i.e., predicting mode shape variation when the structure is subjected to uncertainty, can provide guidance for robust design and control. Nevertheless, computational efficiency is a challenging issue. Direct Monte Carlo simulation is unlikely to be feasible especially for a complex structure with large number of degrees of freedom. In this research, we develop a new probabilistic framework built upon Gaussian process meta-modeling architecture to analyze mode shape variation. To expedite the generation of input dataset for meta-model establishment, a multi-level strategy is adopted which can blend a large amount of low-fidelity data acquired from order-reduced analysis with a small amount of high-fidelity data produced by high-dimensional full finite element analysis. To take advantage of the intrinsic relation of spatial distribution of mode shape, a multi-response strategy is incorporated to predict mode shape variation at different locations simultaneously. These yield a multi-level, multi-response Gaussian process that can efficiently and accurately quantify the effect of structural uncertainty to mode shape variation. Comprehensive case studies are carried out for demonstration and validation.

## 1. Introduction

Mode shapes can be acquired from modal testing experiments on an actual structure or from numerical simulation of its finite element model. The mode shape information is one of the most fundamental properties of a structure, as it essentially decides the spatial pattern of structural vibratory response. Real structures, meanwhile, are inevitably subjected to uncertainties caused by material imperfection, manufacturing tolerance and in-service degradation etc (Liao and Wu, 2018). The deterministic analysis of nominal model without considering uncertainties may render the subsequent design or control ineffective (Zhou and Tang, 2018). Incorporating uncertainties into dynamic modeling and analysis has obvious significance. Intuitively, prediction of dynamic response variation of a structure can be conducted through direct Monte Carlo simulation under given uncertainty parameters. However, for a complicated structure, the number of DOFs (degrees of freedom) in its finite element model is large, leading to high computational cost in solving the eigenvalue problem. When a single run of finite element simulation is computationally demanding, conducting repeated analyses to facilitate direct Monte Carlo simulation becomes infeasible (Yang et al, 2017).

---

[†] Corresponding author

In recent years, there have been continuous efforts in uncertainty quantification of structural dynamic responses. One class of methods aim at reducing the computational time needed for single run through model order reduction. Indeed, along with the advancement of finite element analysis, model order reduction has been one important research subject in computational mechanics/dynamics. A simple and famous approach is referred to as Guyan reduction, where the DOFs in a structure are divided into master DOFs and slave DOFs (Craig and Kurdila, 2006). The effects of the slave DOFs are transformed onto the master DOFs through static condensation, thereby eliminating the slave DOFs in the original model. Salvini and Vivio (2007) applied Guyan reduction into modal analysis at high frequencies. Panayirci et al (2011) utilized it directly to facilitate stochastic structural analysis. To improve the modeling accuracy over Guyan reduction, a variety of component mode synthesis (CMS) approaches have been developed to produce order-reduced models. The fundamental idea of CMS is to retrieve, at least in part, the dynamic effects of truncated DOFs into the order-reduced model. Masson et al (2006) developed a CMS-based model reduction transformation that can be used throughout the entire optimization process to enhance computational efficiency. Shanmugam and Padmanabhan (2006) developed a fixed- and free-interface hybrid CMS method to accurately predict the whirl frequencies of rotor dynamic systems. Zhou and Tang (2016) adopted a NURBS finite element-based free-interface CMS to conduct robust geometry design. While these order-reduction approaches have shown certain effectiveness in mitigating the computational cost of single run, the subsequent sampling-based statistical analysis using direct Monte Carlo simulation still poses significant challenge because it normally requires very large sample size. Besides, it is generally difficult to guarantee the accuracy of results, due to the error introduced by model order reduction.

A different way of realizing uncertainty quantification is through enhancing the efficiency of statistical sampling by means of meta-modeling. A meta-model, once established, can directly predict responses/variations of a process upon given uncertainty parameters without going through Monte Carlo type simulation with large sample size. The establishment or training of meta-model involves a significantly reduced sample size (i.e., the size of dataset containing concerned responses under sampled uncertainty parameters), which leads to greatly reduced computational cost. Amongst various meta-modeling techniques, the Gaussian process architecture exhibits several important advantages (Kennedy and O'Hagan, 2000; O'Hagan, 2006; Rasmussen, 2006). The underlying idea of Gaussian processes is to extend the multivariate Gaussian distribution from a finite dimensional space to an infinite dimensional space. It yields a probabilistic framework for nonparametric regression, thereby addressing the issue of prohibitive sample size required in Monte Carlo simulation. In recent years, there have been attempts of extending Gaussian process into structural dynamic analysis. DiazDelao and Adhikari (2010) employed Gaussian process as an emulator to approximate the frequency response function (FRF) of simple structures, and later (2011) applied Gaussian process together with polynomial chaos expansion to reduce the computational cost of stochastic finite element analyses. Xia and Tang (2013) computed the frequency response values at a small number of frequency points and then built Gaussian process meta-model, based on these values, to predict frequency responses with high resolution of frequency points. Wan et al (2014) proposed a general framework for uncertainty quantification of natural frequencies utilizing Gaussian process meta-model. Zhou and Tang (2017) used Gaussian process to emulate the response surface (i.e., objective function) with respect to design variables in vibration analysis of periodic structure with uncertainty, and carried out robust geometry design to mitigate vibration localization. It is worth noting that the applications mentioned in these studies generally focus on scalar-type response of concern. For example, while frequency response function naturally represents a series of responses at different frequency points, only the responses at certain frequency points are investigated and each response of concern will need a separate Gaussian meta-model. This not only leads to large computational effort if a large number of responses are of interest, but also, overlooks the intrinsic relation of these responses.

For a typical frequency response function, the responses at multiple frequency points are correlated. Similarly, for a give mode shape of a structure, the amplitudes at different DOFs (locations) are correlated intrinsically. Moreover, mode shapes characterize the spatial distribution of responses at their corresponding natural frequencies and, in many cases, it is the distribution pattern, rather than amplitude at specific DOF for a mode, that is of interest. Therefore, if one wants to carry out uncertainty quantification of mode shapes, single-response Gaussian process mentioned above is obviously not an ideal approach. In the realm of statistical meta-modeling, multi-response Gaussian process (MRGP), which is capable of providing emulation of multiple and correlated responses concurrently, has seen recent progresses. Arendt et al (2012a) pointed out that the MRGP could be used to infer true responses when multiple responses were mutually associated with the same input parameters. Wei et al (2018) employed MRGP to produce meta-models for the failure surfaces of system which are then utilized in reliability-based robust design. Bostanabad et al (2018) utilized MRGP for uncertainty quantification analysis of woven fiber composites across multiple scales. Ramirez et al (2018) implemented MRGP to conduct the identification of the ship dynamics and further facilitate the system modeling. Pan et al (2019) built a MRGP meta-model to efficiently characterize the frequency responses with inherent correlation under uncertainties. These investigations have illustrated the possibility of formulating MRGP based meta-modeling to tackle the uncertainty quantification of mode shape information as vectors.

Both the quality and the quantity of training datasets are important in establishing Gaussian process metal-model with high accuracy. In structural dynamic analysis, high-fidelity data can be acquired from experimental measurement or full-scale finite element analysis, the size of which is usually very limited. If one only uses small amount of high-fidelity data as the training dataset, the desired performance of meta-model cannot be ensured. According to O'Hagan and Kennedy (2000), blending a small amount of high-fidelity data with a large amount of low-fidelity data is a promising path. It is worth mentioning that in structural dynamic analysis, those aforementioned order-reduced models derived through such as Guyan reduction or CMS techniques can be utilized to generate large amount of response predictions directly. These responses, containing possible order-reduction error, are naturally low-fidelity data. Indeed, in a recent study (Zhou and Tang, 2018), a multi-level Gaussian process (MLGP) meta-model was established to investigate the variation of single response of a vibration system such as natural frequency. With the large amount of low-fidelity data from order-reduced model, the Gaussian process may avoid those errors associated with the inference procedure. Meanwhile, with the introduction of a few high-fidelity data from full-scale finite element model, one may correct the error of the low-fidelity data inherited from the order-reduction procedure.

The objective of this research is to develop an efficient tool for the uncertainty quantification of mode shape variation. We specifically investigate the development of Gaussian process based meta-model. As indicated, mode shape information is a distributed quantity, requiring a multi-response Gaussian process (MRGP). Meanwhile, we explore the feasibility of incorporating multi-level Gaussian process (MLGP) that can take advantage of the order-reduced modeling techniques available for computational dynamic analysis. This will yield a multi-level, multi-response Gaussian process (MLMRGP) that can adequately address the uncertainty quantification of mode shape information. The rest of the paper is organized as follows. Section 2 first explains the high-fidelity and low-fidelity models and the corresponding datasets to be produced and then used in the new framework. Without loss of generality, order-reduced modeling based on Guyan reduction and CMS is outlined. Subsequently, the mathematical formulation of the proposed MLMRGP is presented in detail. In this framework, we train low-level Gaussian process emulator using low-fidelity datasets produced from order-reduced model, and then train high-level

Gaussian process emulator to further minimize the residual between high- and low-fidelity outputs. The finally established meta-model with hyper-parameters optimized through low- and high-level emulator trainings is used to predict the output given certain input. In the meantime, the output correlations identified in low- and high-level emulators will be assembled to characterize the actual correlation of unseen/testing outputs. In Section 3, comprehensive case studies on a benchmark plate structure are presented, where the effectiveness of the new framework are demonstrated. Concluding remarks are summarized in Section 4.

## 2. Multi-Level Multi-Response Gaussian Process (MLMRGP) for Computational Modal Analysis
### 2.1 Baseline model and two-level datasets

We start from a full-scale finite element model of a vibration system given below,

$$\mathbf{M}\ddot{\mathbf{Z}} + \mathbf{C}\dot{\mathbf{Z}} + \mathbf{KZ} = \mathbf{f} \tag{1}$$

where $\mathbf{M}$, $\mathbf{C}$, and $\mathbf{K}$ are $N \times N$ mass, damping, and stiffness matrices where $N$ is the total number of DOFs, and $\mathbf{Z}$ and $\mathbf{f}$ are the $N$-dimensional displacement response vector and external force vector, respectively. In this research, we study the effects of structural uncertainty to mode shape information, and let $\mathbf{M}$ and $\mathbf{K}$ be functions of $\boldsymbol{\theta}$ where $\boldsymbol{\theta}$ represents the set of uncertain parameters. That is, under uncertainty effect, the mass and stiffness matrices are denoted as $\mathbf{M}(\boldsymbol{\theta})$ and $\mathbf{K}(\boldsymbol{\theta})$. We assume small damping or proportional damping. Therefore, the natural frequencies and mode shapes of the system are determined by the following eigenvalue problem,

$$[\mathbf{K}(\boldsymbol{\theta}) - \omega^2 \mathbf{M}(\boldsymbol{\theta})]\boldsymbol{\psi} = \mathbf{0} \tag{2}$$

Apparently, the natural frequency $\omega$ and the mode shape $\boldsymbol{\psi}$ are affected by $\boldsymbol{\theta}$. Equations (1) and (2) are referred to as the baseline model hereafter, corresponding to the full-scale finite element mesh. The goal of this research is to formulate an efficient and accurate framework from which we can predict the variation of mode shape information. Specifically, we plan to incorporate a two-level Gaussian process approach that can balance between computational cost related to training data acquirement and meta-model accuracy. High-fidelity data can be produced from Equation (2) (i.e., baseline model) directly under sampled uncertainty parameter set $\boldsymbol{\theta}$. In actual practice, the amount of high-fidelity data is usually limited, due to computational cost involved in full-scale finite element analysis.

In order to reduce the overall computational cost, the two-level Gaussian process, to be detailed in the next sub-section, will be trained through the concurrent usage of (a small amount of) high-fidelity data and (a large amount of) low-fidelity data. As indicated in Introduction, in computational structural dynamics, low-fidelity data can be produced using order-reduced methods which however may be subjected to various truncation errors. Through order-reduced analysis that can provide large amount of training data (i.e., mode shape information under sampled uncertainty parameter set), the Gaussian processes emulator can avoid the errors associated with the inference procedure. Meanwhile, with the introduction of a small amount of high-fidelity data based on Equation (2), we can anticipate to correct the error of the low-fidelity data inherited from the order-reduction procedure. Without loss of generality, in this paper we present two representative order-reduced algorithms commonly adopted in structural dynamic analysis, i.e., the Guyan reduction and the fixed-interface CMS algorithms. The formulation details of these two methods are outlined in Appendix for the completeness of the presentation. In Guyan reduction, the DOFs in the baseline full-scale finite element model are divided into master DOFs and slave DOFs. By means of a coordinate transformation based on static condensation, the slave DOFs are eliminated, thereby reducing the dimension of eigenvalue problem analysis. A variety of CMS techniques have been developed in the past. In almost all these techniques, the coordinate transform step is improved, aiming at retrieving the dynamic effects of those DOFs that are eliminated. In a typical CMS procedure, the original, large-scale structure is decomposed into a collection of substructures first, and smaller-size eigenvalue problems are computed for all these separate substructures. Then, a global, order-reduced model is synthesized by combining the reduced-order representations of substructures together with the interface compatibility condition. In the fixed-interface CMS outlined in Appendix, the substructure eigenvalue problems all feature fixed-interface DOFs. The order-reduced model retrieves the interface DOFs (belonging to neighboring substructures) by making the displacements and internal forces compatible at these DOFs. While the details are outlined in Appendix, an order-reduced model can be generically represented as,

$$\bar{\mathbf{M}}\ddot{\mathbf{z}} + \bar{\mathbf{C}}\dot{\mathbf{z}} + \bar{\mathbf{K}}\mathbf{z} = \mathbf{f}_r \quad (3)$$

where $\bar{\mathbf{M}}$, $\bar{\mathbf{C}}$, and $\bar{\mathbf{K}}$ are $m \times m$ mass, damping, and stiffness matrices where $m$ is the total number of DOFs in the order-reduced model, and $\mathbf{z}$ and $\mathbf{f}_r$ are the $m$-dimensional displacement vector and external force vector, respectively. Under uncertainty effect, the eigenvalue problem becomes

$$[\bar{\mathbf{K}}(\boldsymbol{\theta}) - \omega_r^2 \bar{\mathbf{M}}(\boldsymbol{\theta})]\boldsymbol{\psi}_r = \mathbf{0} \quad (4)$$

where $\omega_r$ and $\boldsymbol{\psi}_r$ denote the natural frequency and the corresponding mode shape solved from the order-reduced model. After a coordination transform, one can obtain the mode shape with respect to the original coordinate system. As the order of the system is reduced significantly, we can expect to produce a large amount of mode shape information under uncertainty efficiently from repeatedly solving the eigenvalue problems (Equation (4)).

In summary, Equations (2) and (4) will be employed to generate, respectively, high-fidelity dataset and low-fidelity dataset which are then used as training data for the subsequent two-level Gaussian process meta-modeling.

## 2.2 Multi-level Multi-response Gaussian process framework

We employ the Gaussian process architecture to establish meta-model. In Gaussian process formulation, an unknown system is denoted as $\mathbf{f}(\mathbf{x})$, where $\mathbf{x}$ is an input vector. Here for uncertainty quantification of mode shape variation, the input vector is the set of uncertainty parameters $\boldsymbol{\theta}$ shown in Equations (2) and (4). The observed value of $\mathbf{f}(\mathbf{x})$, i.e., the training data of response, is denoted as $\mathbf{y}$. It is worth noting that, as we are interested in mode shape variation, $\mathbf{y}$ represents the specific mode shape of interest. Therefore $f(\mathbf{x})$ and $\mathbf{y}$ are both vectors, which is the basis of multi-response Gaussian process (MRGP). In the context of computational analysis, we neglect the noise effect, which then results in $\mathbf{f}(\mathbf{x}) = \mathbf{y}$. Given a set of $n_s$ observations described as $\vartheta = \{(\mathbf{y}_i, \mathbf{x}_i), i = 1, 2, \ldots n_s\}$ ($n_s$ is the number of training data), a single-level Gaussian process regression can be implemented to predict the output over target input. Each input $\mathbf{x}_i$ and output $\mathbf{y}_i$ are $r$-dimensional and $q$-dimensional vectors, respectively. In this research, we formulate a two-level MRGP with two types of datasets, i.e., low- and high-fidelity datasets that are introduced as $\vartheta^{(u)} = \{(\mathbf{y}_i^{(u)}, \mathbf{x}_i^{(u)}), i = 1, 2, \ldots n_s^{(u)}; u = 1, 2\}$. The superscript $u$ indicates the fidelity level of data, and each $(\mathbf{y}_i^{(u)}, \mathbf{x}_i^{(u)})$ is referred to as a data point at the $u$-th level. We predict the output vector at target input given two observed datasets $\vartheta^{(1)}$ and $\vartheta^{(2)}$. Specifically, $\vartheta^{(1)}$ is the low-fidelity dataset acquired from order-reduced model (Equation (4)), and $\vartheta^{(2)}$ is the high-fidelity dataset acquired from full-scale baseline model (Equation (2)). Essentially, we will establish a multi-level multi-response Gaussian process (MLMRGP) to facilitate the uncertainty quantification of modal information.

We let the outputs $\mathbf{y}$ in different datasets be expressed, under assumed quasi-linear relations, as

$$\mathbf{Y}^{(1)} = \boldsymbol{\delta}^{(1)}, \qquad \mathbf{Y}^{(2)} = \rho^{(1)}\boldsymbol{\delta}^{(1)} + \boldsymbol{\delta}^{(2)} \qquad (5a, b)$$

where $\rho^{(1)}$ is a regression parameter. $\boldsymbol{\delta}^{(1)}$ and $\boldsymbol{\delta}^{(2)}$ are modeled as two $q$-dimensional independent stationary multivariate Gaussian processes (Zhou and Tang, 2018). As the summation of independent Gaussians remains in the closed form, we can derive the Gaussian process representation of observed low- and high-fidelity data points as

$$\begin{bmatrix} \mathbf{Y}^{(1)} \\ \mathbf{Y}^{(2)} \end{bmatrix} \sim \mathrm{GP}(\mathbf{H}(\mathbf{X})\boldsymbol{\beta}, \mathbf{Q}\boldsymbol{\Sigma}(\mathbf{X}, \mathbf{X}')) \qquad (6)$$

The first item at the right-side of Equation (6), $\mathbf{H}(\mathbf{X})\boldsymbol{\beta}$, represents the linear mean functions of all outputs where $\mathbf{H}(\mathbf{X}) = \begin{bmatrix} \mathbf{h}(\mathbf{X}^{(1)}) & \mathbf{0} \\ \rho^{(1)}\mathbf{h}(\mathbf{X}^{(2)}) & \mathbf{h}(\mathbf{X}^{(2)}) \end{bmatrix}$. Here $\mathbf{X}$ denotes the samples of both low- and high-fidelity inputs.

We use $\mathbf{h}(\mathbf{X}^{(u)}) = \begin{bmatrix} 1 & x_{1,1}^{(u)} & ... & x_{1,r}^{(u)} \\ 1 & x_{2,1}^{(u)} & ... & x_{2,r}^{(u)} \\ .. & .. & .. & .. \\ 1 & x_{n_s^{(u)},1}^{(u)} & ... & x_{n_s^{(u)},r}^{(u)} \end{bmatrix}$ to capture the linear characteristic under small uncertainties, where subscript $n_s^{(1)}$ and $n_s^{(2)}$ are the numbers of low- and high-fidelity datasets, respectively. Therefore the dimension of $\mathbf{H}(\mathbf{X})$ is $(n_s^{(1)} + n_s^{(2)}) \times (2r+2)$. In reality, $n_s^{(2)}$ is much smaller than $n_s^{(1)}$ due to the costly acquisition of high-fidelity datasets through full-scale finite element simulation. $\boldsymbol{\beta}$ is unknown regression coefficient matrix with dimension $(2r+2) \times q$, $\mathbf{Q}$ is non-spatial $q \times q$ matrix representing the covariance among output variables, and $\boldsymbol{\Sigma}$ is a spatial covariance matrix formed by the spatial inputs with dimension $(n_s^{(1)} + n_s^{(2)}) \times (n_s^{(1)} + n_s^{(2)})$. The above equation can be re-organized into vector representation form as shown below,

$$\begin{bmatrix} \text{vec}(\mathbf{Y}^{(1)}) \\ \text{vec}(\mathbf{Y}^{(2)}) \end{bmatrix} = \text{GP}(\mathbf{A}, \mathbf{B}) \tag{7}$$

$$\mathbf{A} = \begin{bmatrix} \text{vec}(\mathbf{h}(\mathbf{X}^{(1)})\boldsymbol{\beta}^{(1)}) \\ \text{vec}(\rho^{(1)}\mathbf{h}(\mathbf{X}^{(1)})\boldsymbol{\beta}^{(1)} + \mathbf{h}(\mathbf{X}^{(2)})\boldsymbol{\beta}^{(2)}) \end{bmatrix} \tag{8a}$$

$$\mathbf{B} = \mathbf{Q} \otimes \begin{bmatrix} \boldsymbol{\Sigma}^{(1)}(\mathbf{X}^{(1)}, \mathbf{X}^{(1)}) & \rho^{(1)}\boldsymbol{\Sigma}^{(1)}(\mathbf{X}^{(1)}, \mathbf{X}^{(2)}) \\ \rho^{(1)}\boldsymbol{\Sigma}^{(1)}(\mathbf{X}^{(1)}, \mathbf{X}^{(2)})^T & \boldsymbol{\Sigma}^{(2)}(\mathbf{X}^{(2)}, \mathbf{X}^{(2)}) + \rho^{(1)2}\boldsymbol{\Sigma}^{(2)}(\mathbf{X}^{(2)}, \mathbf{X}^{(2)}) \end{bmatrix} \tag{8b}$$

where vec(.) is the vectorization operation and $\otimes$ is the Kronecker product. Each entry of the spatial covariance matrix, $\Sigma_{ij}^{(u)} = \exp\left\{-\sum_{k=1}^{r} b_k^{(u)} \left(x_{i,k} - x_{j,k}\right)^2\right\}$, is the so-called squared exponential covariance kernel. Its physical nature enables producing similar outputs when inputs are spatially close, which aligns with our basic understanding (Teimouri et al, 2017). More specifically, we adopt an anisotropic form of this kernel (Rasmussen, 2006), in which the reciprocal of scale-length $b$ is set differently for different inputs. This configuration adaptively adjusts the weights of inputs with respects to associated outputs through optimization, which results in a more accurate meta-model as compared to that built upon the kernel with isotropic form. The anisotropic form generally is used for capturing complex data features, which however renders the optimization computationally intensive as the number of design variables for optimization increases.

The hyper-parameters in the MLMRGP, which include the regression coefficient $\rho^{(1)}$ and the reciprocal scale-length $b_k^{(u)}$ in the covariance kernel, will be identified through learning from the training datasets. Since $\mathbf{Q}$ and $\boldsymbol{\beta}$ depend on these hyper-parameters, only the hyper-parameters denoted as $\phi = [b_1^{(1)}, b_2^{(1)}, ..., b_r^{(1)}, b_1^{(2)}, b_2^{(2)}, ..., b_r^{(2)}, \rho^{(1)}]^T$ need to be optimized. The number of parameters $b_k$ does not have to be equal to input dimension $r$, and generally can be adjusted to accommodate computational capacity. The selection of kernel and the related hyper-parameters relies on the essence of the data used, which can be examined through the cross-validation procedure (Cawley and Talbot, 2010). The key step here is to optimize hyper-parameters following the Bayesian formula,

$$p(\mathbf{Y}^{(2)*} | \mathbf{X}^{(1)}, \mathbf{Y}^{(1)}, \mathbf{X}^{(2)}, \mathbf{Y}^{(2)}, \mathbf{X}^{(2)*}, \phi) = \frac{p(\mathbf{Y}^{(2)*} | \mathbf{X}^{(2)*}, \phi) p(\mathbf{Y}^{(1)}, \mathbf{Y}^{(2)} | \mathbf{X}^{(1)}, \mathbf{X}^{(2)}, \mathbf{Y}^{(2)*}, \phi)}{p(\mathbf{Y}^{(1)}, \mathbf{Y}^{(2)} | \mathbf{X}^{(1)}, \mathbf{X}^{(2)}, \phi)} \quad (9)$$

The solution is referred to as the maximum likelihood estimation (MLE), maximizing the marginal likelihood $p(\mathbf{Y}^{(1)}, \mathbf{Y}^{(2)} | \mathbf{X}^{(1)}, \mathbf{X}^{(2)}, \phi)$ that quantifies the difference between the model prediction under certain hyper-parameters and the corresponding training outputs given the same inputs (Rasmussen, 2006). The likelihood can be further written as the product of the likelihoods at two different levels of emulations (Kennedy and O'Hagan, 2013),

$$p(\mathbf{Y}^{(1)}, \mathbf{Y}^{(2)} | \mathbf{X}^{(1)}, \mathbf{X}^{(2)}, \phi) = p(\mathbf{Y}^{(1)} | \mathbf{X}^{(1)}, b^{(1)}, \sigma_f^{(1)}) p(\mathbf{y}^{(2)} | \mathbf{Y}^{(1)}, \mathbf{X}^{(2)}, b^{(2)}, \sigma_f^{(2)}, \rho^{(1)}) \quad (10)$$

The second term at the right hand side of Equation (10) stands for the likelihood of high-level emulator $\boldsymbol{\delta}^{(2)}$ trained with the high-fidelity dataset utilized to offset the residual error of low-level emulator $\boldsymbol{\delta}^{(1)}$ trained with low-fidelity dataset. Therefore, this term may be re-written as $p(\mathbf{Y}^{(2)} - \rho^{(1)} \mathbf{Y}^{(1),2} |, \mathbf{X}^{(2)}, b^{(2)}, \sigma_f^{(2)}, \rho^{(1)})$ because $\boldsymbol{\delta}^{(2)} = \mathbf{Y}^{(2)} - \rho^{(1)} \mathbf{Y}^{(1)}$ as indicated in Equation (5). Here $\mathbf{Y}^{(1),2}$ denotes the low-fidelity output corresponding to the high-fidelity output $\mathbf{Y}^{(2)}$ under the same input. In other words, the inputs of high-fidelity datasets are a subset of inputs of low fidelity datasets, i.e., $\mathbf{X}^{(2)} \subseteq \mathbf{X}^{(1)}$. Based on the independence condition, we can estimate the parameters $(b^{(1)}, \sigma_f^{(1)})$ that are independent of $(b^{(2)}, \sigma_f^{(2)}, \rho^{(1)})$, by maximizing the logarithms of the aforementioned terms (Johnson and Wichern, 2007),

$$\ln(p(\mathbf{Y}^{(1)} | \mathbf{X}^{(1)}, b^{(1)}, \sigma_f^{(1)}))$$
$$= -\frac{n_s^{(1)}}{2} \ln(\det \mathbf{Q}^{(1)}) - \frac{q}{2} \ln(\det \boldsymbol{\Sigma}^{(1)}) - \frac{1}{2} \text{vec}(\mathbf{Y}^{(1)} - \mathbf{h}(\mathbf{X}^{(1)}) \boldsymbol{\beta}^{(1)})^T (\mathbf{Q}^{(1)} \otimes \boldsymbol{\Sigma}^{(1)})^{-1} \text{vec}(\mathbf{Y}^{(1)} - \mathbf{h}(\mathbf{X}^{(1)}) \boldsymbol{\beta}^{(1)}) \quad (11a)$$

$$\ln(p(\mathbf{Y}^{(2)} | \mathbf{Y}^{(1)}, \mathbf{X}^{(1)}, \mathbf{X}^{(2)}, b^{(2)}, \sigma_f^{(2)}, \rho^{(1)}))$$
$$= -\frac{n_s^{(2)}}{2} \ln(\det \mathbf{Q}^{(2)}) - \frac{q}{2} \ln(\det \boldsymbol{\Sigma}^{(2)}) \quad (11b)$$
$$-\frac{1}{2} \text{vec}(\mathbf{Y}^{(2)} - \mathbf{h}(\mathbf{X}^{(2)}) \boldsymbol{\beta}^{(2)} - \rho^{(1)} \mathbf{Y}^{(1),2})^T (\mathbf{Q}^{(1)} \otimes \boldsymbol{\Sigma}^{(2)})^{-1} \text{vec}(\mathbf{Y}^{(2)} - \mathbf{h}(\mathbf{X}^{(2)}) \boldsymbol{\beta}^{(2)} - \rho^{(1)} \mathbf{Y}^{(1),2})$$

where

$$\boldsymbol{\beta}^{(u)} = [\mathbf{h}(\mathbf{X}^{(u)})^T \boldsymbol{\Sigma}^{(u)} \mathbf{h}(\mathbf{X}^{(u)})]^{-1} \mathbf{h}(\mathbf{X}^{(u)})^T \boldsymbol{\Sigma}^{(u)} \mathbf{Y} \quad (12a)$$

$$\mathbf{Q}^{(u)} = \frac{1}{n_s^{(u)}}(\mathbf{Y} - \mathbf{h}(\mathbf{X}^{(u)})\boldsymbol{\beta}^{(u)})^T \boldsymbol{\Sigma}^{(u)-1}(\mathbf{Y} - \mathbf{h}(\mathbf{X}^{(u)})\boldsymbol{\beta}^{(u)}) \quad (12b)$$

$\mathbf{Y}$ represents $\mathbf{Y}^{(1)}$, if $u=1$; Otherwise $\mathbf{Y}$ represents $\mathbf{Y}^{(2)} - \rho^{(1)}\mathbf{Y}^{(1),2}$. $\boldsymbol{\beta}^{(u)}$ and $\mathbf{Q}^{(u)}$ denote, respectively, the unknown regression coefficient matrix of the mean function and the output covariance matrix involved in the $u$-level emulator.

A sequential, two-step optimization scheme, to be further discussed in the subsequent sub-section, is adopted to identify the optimal hyper-parameters. Once the hyper-parameters $\hat{\boldsymbol{\phi}} = [\hat{b}_1^{(1)}, \hat{b}_2^{(1)}, ..., \hat{b}_k^{(1)}, \hat{b}_1^{(2)}, \hat{b}_2^{(2)}, ..., \hat{b}_k^{(2)}, \hat{\rho}^{(1)}]^T$ and the associated $\hat{\boldsymbol{\beta}}^{(u)}$ and $\hat{\mathbf{Q}}^{(u)}$ are optimized, the target output $\mathbf{Y}^{(2)*}$ over target input $\mathbf{X}^{(2)*}$ can be simply characterized as the posterior Gaussian distribution,

$$\left[\mathbf{Y}^{(2)*}\right] \sim GP(\hat{\boldsymbol{\mu}}(\mathbf{X}^{(2)*}), \hat{\boldsymbol{\Xi}}(\mathbf{X}^{(2)*}, \mathbf{X}^{(2)*'})) \quad (13)$$

The updated mean and covariance functions are given as,

$$\hat{\boldsymbol{\mu}}(\mathbf{X}^{(2)*}) = \text{vec}(\mathbf{H}'\hat{\boldsymbol{\beta}} + \boldsymbol{\Sigma}^{*T}\boldsymbol{\Sigma}^{-1}(\mathbf{Y} - \mathbf{H}\boldsymbol{\beta})) \quad (14a)$$

$$\begin{aligned}\hat{\boldsymbol{\Xi}}(\mathbf{X}^{(2)*}, \mathbf{X}^{(2)*}) &= \hat{\mathbf{Q}} \otimes (\boldsymbol{\Sigma}^{(2)}(\mathbf{X}^{(2)*}, \mathbf{X}^{(2)*}) + \rho^{(1)2}\boldsymbol{\Sigma}^{(1)}(\mathbf{X}^{(2)*}, \mathbf{X}^{(2)*}) - \boldsymbol{\Sigma}^{*T}\boldsymbol{\Sigma}^{-1}\boldsymbol{\Sigma} \\ &+ (\mathbf{H}^* - \boldsymbol{\Sigma}^{*T}\boldsymbol{\Sigma}^{-1}\mathbf{H})\hat{\boldsymbol{\beta}}(\mathbf{H}^* - \boldsymbol{\Sigma}^{*T}\boldsymbol{\Sigma}^{-1}\mathbf{H})^T)\end{aligned} \quad (14b)$$

where

$$\mathbf{H}^* = (\rho^{(1)}\mathbf{h}(\mathbf{X}^{(1)*}), \mathbf{h}(\mathbf{X}^{(2)*}))$$

$$\boldsymbol{\Sigma} = \begin{bmatrix} \boldsymbol{\Sigma}^{(1)}(\mathbf{X}^{(1)}, \mathbf{X}^{(1)}) & \rho^{(1)}\boldsymbol{\Sigma}^{(1)}(\mathbf{X}^{(1)}, \mathbf{X}^{(2)}) \\ \rho^{(1)}\boldsymbol{\Sigma}^{(1)}(\mathbf{X}^{(2)}, \mathbf{X}^{(1)}) & \rho^{(1)2}\boldsymbol{\Sigma}^{(1)}(\mathbf{X}^{(2)}, \mathbf{X}^{(2)}) + \boldsymbol{\Sigma}^{(2)}(\mathbf{X}^{(2)}, \mathbf{X}^{(2)}) \end{bmatrix},$$

$$\boldsymbol{\Sigma}^* = \begin{bmatrix} \rho^{(1)}\boldsymbol{\Sigma}^{(1)}(\mathbf{X}^{(1)}, \mathbf{X}^{(2)*}) \\ \rho^{(1)2}\boldsymbol{\Sigma}^{(1)}(\mathbf{X}^{(2)*}, \mathbf{X}^{(2)}) + \boldsymbol{\Sigma}^{(2)}(\mathbf{X}^{(2)*}, \mathbf{X}^{(2)}) \end{bmatrix}$$

$$\hat{\boldsymbol{\beta}} = [\mathbf{H}^T\boldsymbol{\Sigma}\mathbf{H}]^{-1}\mathbf{H}^T\boldsymbol{\Sigma}\mathbf{Y}, \quad \hat{\mathbf{Q}} = \frac{1}{n_s^{(1)} + n_s^{(2)}}(\mathbf{Y} - \mathbf{H}\boldsymbol{\beta})^T\boldsymbol{\Sigma}^{-1}(\mathbf{Y} - \mathbf{H}\hat{\boldsymbol{\beta}}), \quad \mathbf{Y} = [\mathbf{Y}^{(1)} \ \mathbf{Y}^{(2)}]^T$$

$\hat{\boldsymbol{\beta}}$ and $\hat{\mathbf{Q}}$ are determined by the hyper-parameters that are optimized through the sequential procedure of two-level emulation shown in Equations (11a) and (11b). Collectively they are used to characterize the posterior GP of outputs over target inputs (Equation (13)).

### 2.3 Computational treatment

Since the objective functions (Equations (11a) and (11b)) cannot be expressed in a closed form with respect to the hyper-parameters, sampling-based optimization approaches are preferred. In this study, two algorithms, i.e., simulated annealing (Cao et al, 2019) and particle swarm (Parsopoulos and Vrahatis, 2010), are examined. It is found that particle swarm outperforms simulated annealing in prediction accuracy. Hence, the particle swarm algorithm is adopted in case analysis. The evaluation of objective functions necessitates the computations of matrix inverse and determinant associated with covariance matrices $\mathbf{\Sigma}^{(u)}$ and $\mathbf{Q}^{(u)}$. In general, this may lead to some numerical issues:

- $\mathbf{\Sigma}^{(u)}$ theoretically is positive definite as long as the reciprocal of $b_k^{(u)}$ is greater than 0 (negative $b_k^{(u)}$ is against the physical nature of this kernel). However, it may be nearly singular or ill-conditioned, when a small value pf $b_k^{(u)}$ is statistically sampled during optimization. Extremely ill-conditioned $\mathbf{\Sigma}^{(u)}$ will cause numerical instability in matrix inversion. Numerical computation is generally subjected to resolution (i.e., the smallest non-zero number). Therefore, the determinant of ill-conditioned $\mathbf{\Sigma}^{(u)}$ cannot be differentiated.
- $\mathbf{Q}^{(u)}$ should be positive definite as well. However, it is often close to being singular, especially when a large number of output variables are involved. Earlier studies have noted that very large number of output variables is not recommended since it may induce numerical instability (Arendt et al, 2012b).
- $\mathbf{Q}^{(u)} \otimes \mathbf{\Sigma}^{(u)}$ yields a high-dimensional matrix when many response variables and training datasets are taken into account. The inversion of such a large matrix required in each iteration of objective function evaluation is computationally expensive.

Our strategies to address these issues are summarized as follows:

- *Matrix inversion:* We monitor the condition numbers of $\mathbf{\Sigma}^{(u)}$ and $\mathbf{Q}^{(u)}$, and set a threshold to decide if current objective evaluation is executed or ignored. Meanwhile, we add a small diagonal perturbation into matrices to be inverted.
- *Matrix determinant:* We incorporate matrix decomposition, i.e., LU decomposition, or eigenvalue analysis to compute the determinant.
- *Large-size matrix inverse:* We take advantage of a Kronecker product principle, i.e., $(\mathbf{Q}^{(u)} \otimes \mathbf{\Sigma}^{(u)})^{-1} = \mathbf{Q}^{(u)-1} \otimes \mathbf{\Sigma}^{(u)-1}$ (Loan, 2000), where $\mathbf{\Sigma}^{(u)}$ and $\mathbf{Q}^{(u)}$ are both invertible. Recall that the computational complexity of inverting a $P \times P$ matrix is $O(P^3)$. The original complexity $O((n_s^{(u)} \times q)^3)$ can be reduced to $O(n_s^{(u)3}) + O(q^3)$.

## 3. Case Studies: Meta-Model Establishment and Uncertainty Quantification Illustration

In this section, we demonstrate the effectiveness of the proposed framework. We specifically focus on the mode shape variations, and utilize the multi-level multi-response Gaussian process approach. We highlight the influences of low-fidelity and high-fidelity datasets to the uncertainty quantification performance.

## 3.1 Benchmark structure and data preparation

*3.1.1 Nominal structure and model order reduction*

We consider a benchmark structure shown in Figure 1(a). It consists of essentially 3 rectangular plates connected together. For the nominal structure without uncertainty, the mass density and Young's modulus are 7850 kg/m$^3$ and 206 GPa. From bottom to top, these three plates have, respectively, 2,214, 630, and 858 DOFs. Altogether, the full-scale finite element model of this benchmark structure has 3,510 DOFs. We choose this structural configuration so interested readers can readily re-construct the mesh for validation and comparison. This structure can be directly decomposed into substructures to facilitate various order-reduction analysis. The order-reduction approach adopted hereafter can be extended easily to more complicated structures where substructure decomposition is straightforward.

As mentioned in the preceding sections, one important component of this proposed methodology is to incorporate low-fidelity datasets into uncertainty quantification, which has the prospect of significantly reducing the computational cost needed for the generation of training data for meta-model establishment. Commonly, Guyan reduction and component mode synthesis (CMS) approaches are used in order-reduction of structural dynamic analysis. The standard Guyan reduction and fixed-interface CMS are outlined in Appendix. In Guyan reduction, the DOFs are first divided into master DOFs and slave DOFs, and the responses of the slave DOFs are transformed onto the master DOFs through static condensation. The fixed-interface CMS takes into consideration the dynamic effects of the DOFs that are truncated in order-reduction, and therefore is generally more accurate at the price of additional computations compared with Guyan reduction. Here we analyze both order-reduction methods. We consider the first three *z*-direction bending modes. For Guyan reduction, the master DOFs selected are indicated in Figure 1(b). Apparently, *z*-direction DOFs that are away from the clamped boundaries play a dominant role in these modes, which are selected as the master DOFs. Specifically, 150, 48 and 32 master DOFs are selected for three substructures (from bottom to top), respectively, yielding an order-reduced model with 230 DOFs in total. For fixed-interface CMS, we keep the first 10, 5, and 2 modes of the substructures (from bottom to top), respectively. We also keep all the interface DOFs (between neighboring substructures) in the order-reduced model. The CMS order-reduced model thus has 209 DOFs in total.

The computation is carried out on a 2-processor desktop (Intel E5620@2.4GHz) under MATLAB environment. In this research, we use self-developed finite element code to carry out the investigations. This will facilitate a streamlined process to generate datasets with multiple fidelity levels. The finite element model of the benchmark structure used in the analysis is fully validated using ANSYS. When the first 20 natural frequencies and mode shapes are sought, the full finite element analysis takes 1.15s to complete one run. In comparison, the Guyan reduction and the fixed interface CMS take 0.13s and 0.25s, respectively. The first 5 natural frequencies are listed in Table 1, and the mode shapes are shown in

Figure 2. For illustration purpose, we present only the z-direction DOFs in mode shape comparison since the modes involved are all bending modes. In general, the natural frequencies and mode shapes obtained from order-reduced approaches have good accuracy as compared with full-scale finite element analysis. The performance degrades as mode order increases. Unsurprisingly, the fixed-interface CMS generally outperforms Guyan reduction in terms of accuracy, while the results are somewhat comparable. Since our goal is to demonstrate that a two-level Gaussian process that integrates together a small amount of high-fidelity data with a large amount of low-fidelity data can yield a satisfying meta-model, in what follows we adopt Guyan reduction as the low-fidelity data generator. The Guyan reduction features faster computation with lower fidelity (i.e., less accuracy), and therefore can better highlight the advantage of two-level meta-modeling.

*3.1.2 Model uncertainties and dataset preparation*

We assume model uncertainties come from material properties, i.e., mass density and Yong's modulus. Specifically, the benchmark structure is divided into 6 segments (as shown in Figure 1), and each segment features its material property uncertainties, leading to 12 uncertainty parameters. We let these 12 uncertainty parameters be subjected to multivariate normal distribution, in which the means take the nominal values and the standard deviations are set as 20% of the means. Following Latin hypercube sampling (Kroese et al, 2011), we generate 1,000 uncertainty input samples. The sampled uncertainty parameters are then employed in Monte Carlo simulations of both full-scale finite element analysis and Guyan reduction. In this case study, since the full-scale finite element mesh of the benchmark structure has relatively low dimension, we can readily produce the Monte Carlo simulation results which are then used for validation.

As mentioned in Section 2.3, the number of output variables of the meta-model may not be very large, in order to yield tractable computation and also to avoid numerical instability. As such, for each vibration mode of interest, we focus on 50 DOFs on the top surface of the structure where the mode shape amplitudes of the nominal structure have the largest absolute values. In other words, these DOFs are used to represent/characterize the respective mode shapes. The multiple responses defined in MLMRGP hence consist of these 50 inter-related mode shape amplitudes.

Figure 3 shows the variations of the first 2 bending modes obtained from the Monte Carlo simulation of full-scale finite element analysis. At each selected DOF, the probabilistic density function (PDF) based on 1,000 uncertainty samples is shown as the violin plot. Similarly, Figure 4 shows the results of the first 2 bending modes obtained from the Monte Carlo simulation of Guyan reduction analysis. There are notable discrepancies when we compare Figures 3 and 4. Apparently, the second mode shape is more sensitive to uncertainties. In addition, the output distribution at each DOF varies. In the subsequent analysis, these mode shape data will be used for meta-model training and validation.

**3.2 MLMRGP meta-model establishment and validation**

*3.2.1 Meta-model establishment*

The multi-level multi-response Gaussian process (MLMRGP) proposed in this research takes advantage of the multi-fidelity datasets generated in Section 3.1.2, and takes into consideration the inherent correlation among mode shape amplitudes. We start from employing 30 high-fidelity data and 300 low-fidelity data, both of which are randomly selected from the respective databases generated by Monte Carlo simulation of full-scale finite element and order-reduced analysis (Section 3.1.2). The remaining 700 low-fidelity data and the corresponding 700 high-fidelity data (under the same uncertainty inputs/parameters) are used for validation. To facilitate the optimization of hyper-parameters through Equations (11a) and (11b), preprocessing of input/output data is necessary. The input data (i.e., the set of uncertainty parameters) are converted into standard normal distributions, and the output data, i.e., mode shape amplitudes are scaled to [-1 1].

In establishing the meta-model using MLMRGP, we adopt linear mean and anisotropic exponential covariance kernels (Equations (6) and 8(c)). The exponential covariance kernel at each level's emulator includes 6 reciprocals of scale-length values. Each scale-length is used to weigh the spatial correlation of two input samples, i.e., the variations of mass density and Young's modulus of one specific segment in the structure analyzed. For example, $b_k^{(u)}$ characterizes the spatial correlation of parameterized inputs, i.e., the mass density and Young's modulus of the *k*-th sector in the *u*-th level emulator. In addition to 6 $b_k^{(u)}$ at each level's emulator, there is one regression coefficient considered. Therefore, a total of 13 hyper-parameters are to be optimized. Using less number of hyper-parameters would render the meta-model incapable of capturing the underlying data features. On the other hand, more hyper-parameters would increase the computational cost and may cause model overfitting. Particle swarm algorithm is used for hyper-parameter optimization. We need to define the design boundaries for all hyper-parameters. Here, 6 $b_k^{(1)}$ and 6 $b_k^{(2)}$ are specified with bounds [0.01, 50] and [0.01, 300], respectively. Regression coefficient $\rho^{(1)}$ is specified with bound [0.001, 1]. The simulation variables, i.e., swarm size and maximum iteration number of particle swarm algorithm are set as 300 and 50,000, respectively. The MLMRGP based meta-model is then established following the procedure outlined in Section 2.2.

*3.2.2 Characterization of the 1$^{st}$ mode shape variation*

Once the meta-model is trained using the MLMRGP framework, we can use it as emulator to predict mode shape variation under given input parameters (i.e., various uncertainty parameters). Recall that 700 high-fidelity and low-fidelity datasets, under the same 700 samples of uncertainty parameters, are not used in training. They will be use in validation. Using these 700 samples of uncertainty parameters, we can predict the corresponding mode shape outputs through the meta-model established. We consider the high-fidelity, full-scale finite element results as the accurate results. The prediction errors of the 1$^{st}$ mode shape amplitudes by the meta-model are shown in Figure 5, where the PDF of mode amplitude at given DOF is estimated based on 700 prediction error values. It is worth noting that the peaks of PDFs do not truly represent the worst case (i.e., maximum) errors. For clear illustration, we include the worst case errors over entire DOFs in the plot, which are marked as crosses (Figure 5). The results show that the mean errors vary slightly at different DOFs. Overall, however, the mean errors are all below 2%. The worst case errors versus DOFs follow the similar trend, and are all under 8%.

To further assess the *overall* error level, we define the Average of Mean Errors (AMR),

$$\text{AMR} = \frac{1}{m}\sum_{i=1}^{m}\overline{e}_i \qquad (15)$$

where $\overline{e}_i$ represents the mean error at the *i*-th DOF (over the 700 samples), and *m* denotes the number of DOFs selected in mode shape characterization. In this case study, $m = 50$. The AMR of the 1$^{st}$ mode shape variation prediction is 1.08%. Recall that the low-fidelity dataset is generated by Guyan reduction and bears order-reduction error. For comparison purpose, we calculate the AMR for the corresponding 700 low-fidelity data directly, and find that for the 1$^{st}$ mode the low-fidelity data as a whole yields an AMR of 1.18%. This indicates that the MLMRGP can effectively improve the prediction accuracy over the original low-fidelity dataset.

We now take further look at some prediction instances. For example, the 26$^{th}$ and the 43$^{rd}$ DOFs show larger prediction errors. Recall that the posterior mean values of MLMRGP are employed as the prediction results here. Meanwhile, the prediction results of MLMRGP are actually statistically characterized by posterior mean and covariance. The posterior covariance essentially indicates the confidence/likelihood of posterior mean. We then analyze the PDFs of prediction results of mode amplitudes at DOFs of interest that are built upon the posterior mean and covariance. The covariance in this case is the variance, as we focus on statistical relation of samples that have the same response variable. For comparison, we also include good prediction instances, i.e., the 42$^{nd}$ DOF and the 48$^{th}$ DOF with smaller prediction errors. We attempt to make the comparison for DOFs with similar nominal mode shape amplitudes. Figure 6 shows the prediction performance comparison from a probabilistic perspective. A shaded area denotes the region between plus and minus one standard deviations. It can be observed that all true values fall within the shaded areas of predicted PDF. While the worst case errors (i.e., deviation between the posterior mean and the actual value) in the top two sub-plots (the 43$^{rd}$ and the 26$^{th}$ DOFs) are much larger, the corresponding variance increases significantly. This indicates that while relatively larger errors occur at these DOFs, the meta-model is capable of pointing out the low confidence at these locations. This capability of probabilistic prediction illustrates that the Gaussian process is a powerful statistical meta-modeling technique.

One important feature of the proposed MLMRGP framework is its capability of taking the correlation of different outputs (mode amplitudes at different DOFs) into consideration. The covariance matrix $\hat{\mathbf{Q}}$ identified (Equation (14b)) reflects the most probable correlation among outputs. In order to facilitate the comparison, this covariance matrix is converted to the correlation matrix in the following manner (Shynk, 2013),

$$\mathbf{D} = \sqrt{\text{diag}(\mathbf{Cov})}, \qquad \mathbf{Corr} = \mathbf{D}^{-1}\mathbf{Cov}\mathbf{D}^{-1} \qquad (16a,b)$$

where **Cov** and **Corr** represent the original covariance matrix and the resultant correlation matrix, respectively. The output correlation is then used to evaluate the MLMRGP framework by comparing it with the true correlation of testing datasets. In this case, it is interesting to observe that the values in correlation matrix are all close to 1, which indicates the high correlation of all response variables. We arbitrarily choose 4 out of 50 response variables for comparison, as shown in Figure 7. The top-left and bottom-right numbers represent the true correlation of testing dataset and the correlation identified from meta-model prediction, respectively. Clearly, they match quite well, illustrating that MLMRGP is capable of accurately identifying statistical correlation among different response variables.

*3.2.3 Characterization of the 2$^{nd}$ mode shape variation*

Following a similar process, we analyze and interpret the meta-model prediction of the 2$^{nd}$ mode shape variation. We first analyze the order-reduction error of the entire low-fidelity testing dataset as a whole, and the AMR (Equation (15)) calculated is 6.91%. Utilizing the MLMRGP meta-model, we carry out emulation and the AMR for GP prediction is calculated as 3.39%. This indicates a significant improvement through the MLMRGP process due to the incorporation of a small amount (30) high-fidelity data. The results are shown in Figure 8. The largest error occurs at the 48$^{th}$ DOF with the mean error at around 9%. The 30$^{th}$, 36$^{th}$ and 37$^{th}$ DOFs also exhibit considerable errors. Recall Figure 3. One may readily notice that the error magnitude is generally associated with the original response variation distribution. The larger the variance of original output distribution is, the larger the corresponding errors will likely be, simply because the variance reflects the sensitivity of mode shape with respect to input uncertainty parameters. Additionally, the low-fidelity dataset of the 2$^{nd}$ mode inherently has greater error than that of the 1$^{st}$ mode shape, which can be seen in the deterministic analysis result (Figure 2). The framework of MLMRGP allows us to tune/optimize the hyper-parameters to ensure prediction accuracy of multiple response variables (i.e., mode shape amplitudes of interest). It reduces large errors at certain DOFs while at the same time it may indeed yield trade-off at some other DOFs. The final prediction errors reflect how the trained meta-model fits the testing datasets. While the facts mentioned above indeed pose a challenge for the mode shape amplitude prediction, the MLMRGP outperforms the Monte Carlo simulation utilizing Guyan-order reduction analysis with much higher accuracy.

Some example prediction instances are examined probabilistically as shown in Figure 9. Once again, all true values are within the region between plus and minus one standard deviations. The result illustrates that large prediction error generally occurs with large variance of predicted PDF, which reflects the confidence level of predicted output. The identified correlation (bottom right number) and the true correlation (upper left number) extracted from testing datasets are put together for comparison in Figure 10. The correlation among different outputs becomes more complicated due to larger sensitivity of the $2^{nd}$ mode shape with respect to uncertainty parameters. The negative correlation here indicates a relationship between two response variables in which one variable increases and the other decreases. It can be observed that the general trend/pattern of correlation is completely captured, which verifies that MLMRGP takes output correlation into consideration during emulation analysis.

*3.2.3 Effect of training dataset size*

An important condition in the formulation of the MLMRGP framework (Section 2.2) is that the inputs to the high-fidelity training dataset are a subset of the inputs to the low-fidelity training dataset. In other words, within this case study setup, when the size of the low-fidelity dataset remains to be the same (i.e., 300), the size of high-fidelity training datasets can be adjusted from 0 to 300. The high-fidelity training data are reliable evidences used to correct the meta-model error owing to the low-fidelity model truncation. Figure 11 shows the AMR results for the first two modes as we increase the high-fidelity training data size. Unsurprisingly, both show performance improvement.

*3.2.4 Meta-model cross-validation*

The effect of increasing size of high-fidelity training data size indicated in the preceding sub-section is intuitive. It is also worth noting that the accuracy improvement may not be simply proportional to the training data size. Essentially, the performance has to do with whether the training data set captures the underlying features of the output variables. The training data set, on the other hand, is generated based on random inputs. When the training dataset changes, one may expect change of the prediction performance. In order to examine how well a meta-model generalizes to new datasets and to avoid model under-fitting or over-fitting, we apply the cross-validation analysis. Particularly, here we use the bootstrap sampling-based cross-validation, which allows the random sampling with replacement (Davison and Hinkley, 2009). We use the same sizes of low-fidelity and high-fidelity datasets, 300 and 30 respectively. 5 emulations with different randomly selected training and testing datasets are implemented. Table 2 shows the AMR values of both mode shapes under different emulations. It is found that the results under different emulations are quite consistent, showing the robustness of MLMRGP. Besides, the mean of AMR values using MLMRGP is always smaller than the AMR of low-fidelity data evaluated as a whole, which demonstrates the effectiveness of MLMRGP.

## 4. Conclusion

A new multi-level, multi-response Gaussian process (MLMRGP) meta-modeling technique is developed in this research, aiming at uncertainty quantification of mode shape variation. This framework allows the usage of a small amount of high-fidelity data produced by full-scale finite element analysis together with a large amount of low-fidelity data produced by order-reduced model such as Guyan reduction as training datasets for meta-model establishment. This reduces significantly the computational cost needed for generating the training data. The new framework also yields the simultaneous prediction of mode shape amplitudes at different DOFs, thereby capturing their intrinsic correlations. Case studies using a benchmark structure indicates that the MLMRGP technique can effectively characterize the mode shape variations. The incorporation of a small amount of high-fidelity data can increase the prediction accuracy compared with using order-reduced data alone. This framework can be extended to general structural dynamic analysis concerning multiple output responses.

## Acknowledgment

This research is supported in part by National Science Foundation under grant CMMI – 1825324.

## Appendix: Order-reduced Models

This section outlines the mathematical formulations of Guyan reduction and fixed-interface component mode synthesis (CMS).

*Guyan reduction*

Guyan reduction is a well-established model order reduction technique, where the DOFs (degrees of freedom) are divided into master and slave DOFs. This division can be expressed in the matrix form as follows (Craig and Kurdila, 2006),

$$\begin{bmatrix} \mathbf{M}_{mm} & \mathbf{M}_{ms} \\ \mathbf{M}_{sm} & \mathbf{M}_{ss} \end{bmatrix} \begin{bmatrix} \ddot{\mathbf{Z}}_m \\ \ddot{\mathbf{Z}}_s \end{bmatrix} + \begin{bmatrix} \mathbf{K}_{mm} & \mathbf{K}_{ms} \\ \mathbf{K}_{sm} & \mathbf{K}_{ss} \end{bmatrix} \begin{bmatrix} \mathbf{Z}_m \\ \mathbf{Z}_s \end{bmatrix} = \begin{bmatrix} \mathbf{0} \\ \mathbf{0} \end{bmatrix} \qquad (A.1)$$

where subscripts *m* and *s* denote the master and slave DOFs, respectively. The second row in the above matrix equation yields

$$\mathbf{Z}_s = -\mathbf{K}_{ss}^{-1}(\mathbf{M}_{sm}\ddot{\mathbf{Z}}_m + \mathbf{M}_{ss}\ddot{\mathbf{Z}}_s + \mathbf{K}_{sm}\mathbf{Z}_m) \tag{A.2}$$

Neglecting the inertia terms in Equation (A.2) results in the transformation matrix $\mathbf{T}_G$ for Guyan reduction,

$$\begin{bmatrix} \mathbf{Z}_m \\ \mathbf{Z}_s \end{bmatrix} = \begin{bmatrix} \mathbf{I} \\ -\mathbf{K}_{ss}^{-1}\mathbf{K}_{sm} \end{bmatrix} \mathbf{Z}_m = \mathbf{T}_G \mathbf{Z}_m \tag{A.3}$$

The order-reduced stiffness and mass matrices can be obtained as

$$\bar{\mathbf{M}} = \mathbf{T}_G^T \mathbf{M} \mathbf{T}_G \tag{A.4a}$$

$$\bar{\mathbf{K}} = \mathbf{T}_G^T \mathbf{K} \mathbf{T}_G \tag{A.4b}$$

*Fixed-interface component mode synthesis*

In component mode synthesis (CMS) based order reduction, a structure is divided into a group of substructures first (Sarsri et al, 2011). For the *s*-th substructure, the DOFs are divided into interior DOFs and interface DOFs (between adjacent substructures). Its equation of motion under free vibration condition can be written as

$$\begin{bmatrix} \mathbf{M}_{ii}^s & \mathbf{M}_{ij}^s \\ \mathbf{M}_{ji}^s & \mathbf{M}_{jj}^s \end{bmatrix} \begin{bmatrix} \ddot{\mathbf{Z}}_i^s \\ \ddot{\mathbf{Z}}_j^s \end{bmatrix} + \begin{bmatrix} \mathbf{K}_{ii}^s & \mathbf{K}_{ji}^s \\ \mathbf{K}_{ji}^s & \mathbf{K}_{jj}^s \end{bmatrix} \begin{bmatrix} \mathbf{Z}_i^s \\ \mathbf{Z}_j^s \end{bmatrix} = \begin{bmatrix} \mathbf{0} \\ \mathbf{f}_j^s \end{bmatrix} \tag{A.5}$$

where subscript *i* and *j* indicate the interior and interface DOFs. $\mathbf{f}_j^s$ represents the internal force due to neighboring structure. In fixed-interface CMS, at the substructure level we let $\mathbf{Z}_j^s = 0$ and subsequently solve the eigenvalue problem,

$$(\mathbf{K}_{ii}^s - \omega_{ii}^{s2}\mathbf{M}_{ii}^s)\mathbf{\psi}_{ii}^s = \mathbf{0} \tag{A6}$$

where $\mathbf{\psi}_{ii}^s$ denotes the eigenvector set of the fixed-interface substructure. In CMS, order-reduction is facilitated by retaining only the lower-order eigenvectors to represent the dynamic characteristics of each substructure. All interface DOFs are retained in the order reduced model. Let $\mathbf{\psi}_k^s$ denote the set of kept eigenvectors of the *s*-th substructure. As a basic fixed-interface CMS, we apply a static condensation to take into account the coupling between interface DOFs and the interior DOFs. The transformation between the original DOFs and the order-reduced DOFs for the *s*-th substructure can then be expressed as

$$\mathbf{T}_{CMS}^s = \begin{bmatrix} \mathbf{\psi}_k^s & \mathbf{\psi}_{ij}^s \\ & \mathbf{I}^s \end{bmatrix} \tag{A.7}$$

where $\boldsymbol{\psi}_{ij}^{s} = -\mathbf{K}_{ii}^{s-1}\mathbf{K}_{ij}^{s}$ and $\mathbf{I}^{s}$ is an identity matrix. Using the above transformation matrix, order-reduced mass and stiffness matrices can be formed. It is worth noting that, while CMS generally yields much improved accuracy as compared with Guyan reduction, eigenvalue analysis at the substructure level is needed which increases computational cost.

**References**

Arendt, P. D., Apley, D. W., Chen, W., Lamb, D., and Gorsich, D., 2012a, "Improving identifiability in model calibration using multiple responses," ASME Journal of Mechanical Design, 134(10), pp. 100909.

Arendt, P. D., Apley, D. W., and Chen, W., 2012b, "Quantification of model uncertainty: calibration, model discrepancy, and identifiability," ASME Journal of Mechanical Design, 134(10), pp. 100908.

Bostanabad, R., Liang, B., Gao, J., Liu, W.K., Cao, J., Zeng, D., Su, X., Xu. H., Li, Y., and Chen, W., 2018, "Uncertainty quantification in multiscale simulation of woven fiber composites," Computer Methods in Applied Mechanical and Engineering, 338, pp. 506-532.

Cao, P., Shuai, Q., and Tang, J., 2019, "Leveraging Gaussian process regression and many-objective optimization through voting scores for fault identification," IEEE Access, 7, 94481-94496.

Cawley, G.C., and Talbot, N.L.C., 2010, "On over-fitting in model selection and subsequent selection bias in performance evaluation," Journal of Machine Learning Research, 11, pp. 2079-2107.

Craig, R.R., and Kurdila, A.J., Fundamentals of Structural Dynamics, Wiley, 2006.

Davision, A.C., and Hinkley, D.V., Bootstrap Methods and Their Applications, Cambridge University Press, 2009.

DiazDelaO, F.A., and Adhikari, S., 2010, "Structural dynamic analysis using Gaussian process emulators," Engineering Computations, 27(5), pp. 580-605.

DiazDelaO, F.A., and Adhikari, S., 2011, "Gaussian process emulators for the stochastic finite element method," International Journal for Numerical Methods in Engineering, 87(6), pp. 521-540.

Johnson, R., and Wichern, D., Applied Multivariate Statistical Analysis, Prentice-Hall, 2007.

Kennedy, M.C., and O'Hagan, A., 2000, "Predicting the output from a complex computer code when fast approximation are available," Biometrika, 87(1), pp. 1-13.

Kroese, D.P., Taimre, T., and Botev, Z.I., Handbook of Monte Carlo Methods, Wiley, 2011.

Parsopoulos, K.E., and Vrahatis, M.N., Particle Swarm Optimization and Intelligence: Advances and Applications, IGI Global, 2010.

Liao, H.T., and Wu, W.W., "A frequency domain method for calculating the failure probability of nonlinear systems with random uncertainty," ASME Journal of Vibration and Acoustics, 140(4), 041019.

Loan, C.F.V., 2000, "The ubiquitous Kronecker product," Journal of Computational and Applied Mathematics, 123(1-2), pp. 85-100.

Masson, G., Ait Brik, B., Cogan, S., and Bouhaddi, N., 2006, "Component mode synthesis (CMS) based on an enriched ritz approach for efficient structural optimization," Journal of Sound and Vibration, 296(4-5), pp. 845-860.

O'Hagan, A., 2006, "Bayesian analysis of computer code outputs: A tutorial," Reliability Engineering and System Safety, 91(10-11), pp. 1290-1300.

Pan, W., Tang, G., and Tang, J., 2019, "Frequency response-based uncertainty analysis of vibration system utilizing multiple response gaussian proess," ASME Journal of Vibration and Acoustics, 141(5), 051010.

Panayirci, H.M., Pradlwater, H.J., and Schueller, G.I., 2011, "Efficient stochastic structural analysis using Guyan reduction," Advances in Engineering Software, 42(4), pp. 187-196.

Rasmussen, C.E., and Williams, C.K.I., Gaussian Process for Machine Learning, MIT press, 2006.

Salvini, P., and Vivio, F., 2007, "Dynamic reduction strategies to extend modal analysis approach at higher frequencies," Finite Elements in Analysis and Design, 43(2), pp. 931-940.

Shanmugam, A., and Padmanabhan, C., 2006, "A fixed-free interface component mode synthesis method for rotordynamic analysis," Journal of Sound and Vibration, 297(3-5), pp. 664-679.

Sarsri, D., Azrar, L., Jebbouri, A., and EI Hami, A., 2011, "Component mode synthesis and polynomial chaos expansions for stochastic frequency functions of large linear FE models," Computers and Structures, 89(3-4), pp. 346-356.

Syhnk, J.J., 2013, Probability, Random Variables, and Random processes: Theory and Signal Processing Applications, John Wiley & Sons.

Teimouri, H., Milani, A.S., Loeppky, J., and Seethaler, R., 2017, "A Gaussian process–based approach to cope with uncertainty in structural health monitoring," Structural Health Monitoring, 16, pp. 174-184.

Wan, H.P., Mao, Z., Todd, M.D., and Ren, W.X., 2014, "Analytical uncertainty quantification for modal frequencies with structural parameter uncertainty using a Gaussian process metamodel," Engineering Structures, 75, pp. 577-589.

Wei, P., Liu, F., and Tang, C., 2018, "Reliability and reliability-based importance analysis of structural systems using multiple response Gaussian process model," Reliability Engineering & System Safety, 175, pp. 183-195.

Xia, Z., and Tang, J., 2013, "Characterization of Dynamic Response of Structures with Uncertainty by Using Gaussian Processes," ASME Journal of Vibration and Acoustics, 135(5), pp. 051006.

Yang, J., Faverjon, B., Peters, H., Marburg, S., and Kessissoglou, N., 2017, "Deterministic and stochastic model order reduction for vibration analyses of structures with uncertainties", ASME Journal of Vibration and Acoustics, 139(2), 021007.

Zhou, K., Hegde, A., Cao, P., and Tang, J., 2017, "Design optimization towards alleviating forced response variation in cyclically periodic structure using Gaussian process," ASME Journal of Vibration and Acoustics, 139(1), 011017.

Zhou, K., Liang, G., and Tang, J., 2016, "Component mode synthesis order-reduction for dynamic analysis of structure modeled with NURBS finite element," ASME Journal of Vibration and Acoustics, 138(2), 021016.

Zhou, K., and Tang, J., 2018, "Uncertainty quantification in structural dynamic analysis using two-level Gaussian processes and Bayesian inference," Journal of Sound and Vibration, 412, pp. 95-115.

Table 1. First 5 natural frequencies of the nominal structure.

| Mode order | Full-scale finite element | Guyan reduction | Fixed-interface CMS |
|:---:|:---:|:---:|:---:|
| 1 | 144.3078 | 144.5716 | 142.5492 |
| 2 | 334.7630 | 345.8223 | 330.8748 |
| 3 | 367.8749 | 373.9271 | 362.5558 |
| 4 | 571.4485 | 596.7220 | 566.7540 |
| 5 | 709.2347 | 828.6293 | 702.0148 |

Table 2. AMR comparion between cross validation of MLMRGP and low-fideltiy testing data

|  | AMR (%) | | |
| --- | --- | --- | --- |
|  | Prediction using MLMRGP built upon 300 low-fidelity data & 30-fidelity data | | Prediction using Guyan order-reduced model |
| $1^{st}$ mode shape | 1.081<br>0.864<br>1.043<br>1.119<br>1.010 | **Mean:1.023**<br><br>STD:0.098 | **1.183** |
| $2^{nd}$ mode shape | 3.390<br>2.911<br>3.057<br>3.467<br>3.609 | **Mean:3.286**<br><br>STD:0.292 | **6.910** |

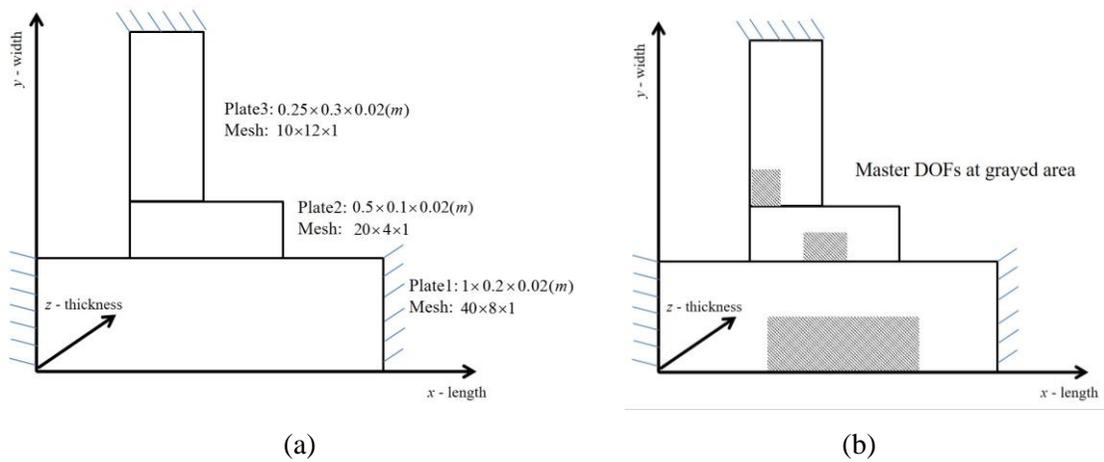

(a)                                       (b)

Figure 1. Benchmark structure. (a) Configuration; (b) DOFs at grayed areas are selected as the master DOFs in Guyan reduction.

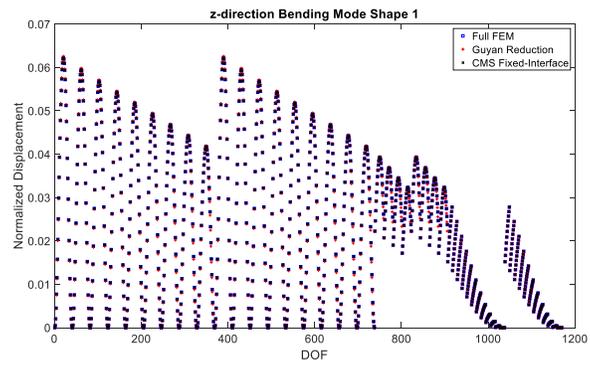

(a)

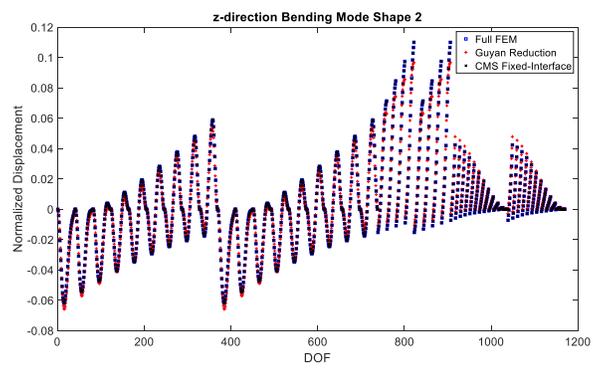

(b)

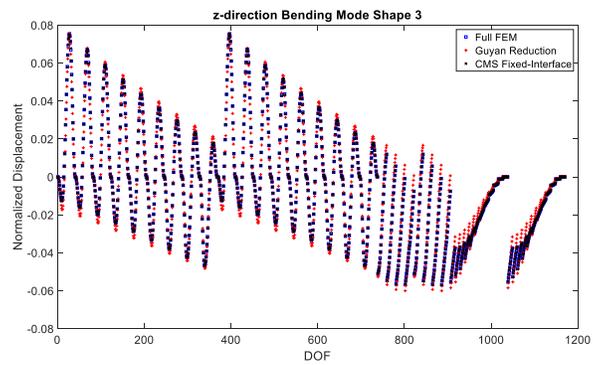

(c)

Figure 2. Mode shape comparison. (a) 1st $z$-direction bending mode shape; (b) 2nd $z$-direction bending mode shape; (c) 3rd $z$-direction bending mode shape.

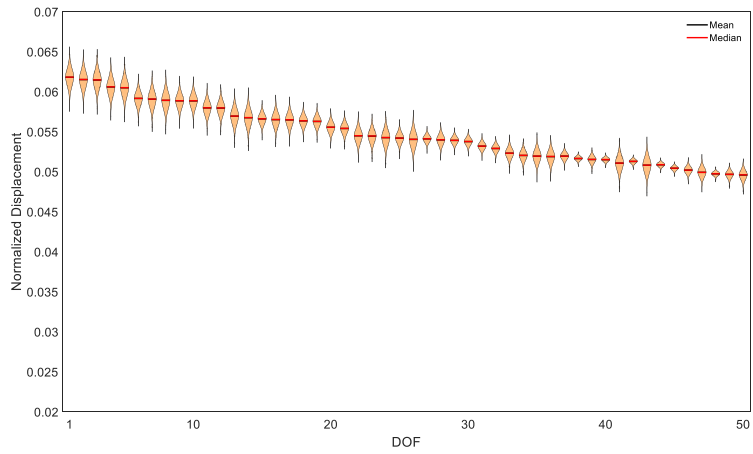

(a)

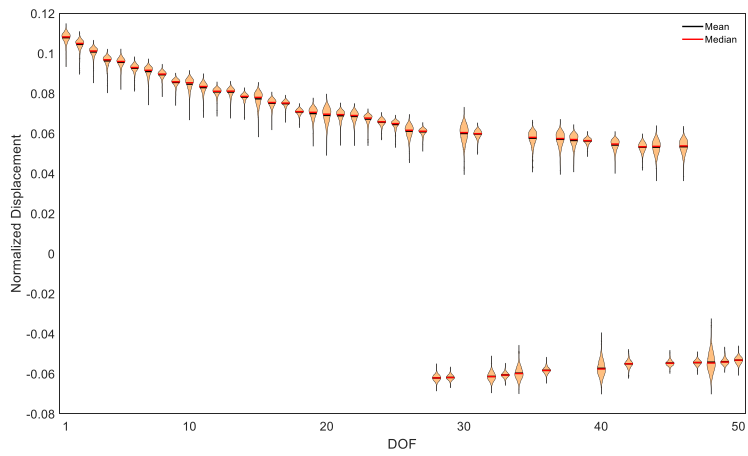

(b)

Figure 3. Distributions of mode shape amplitudes at selected DOFs using full-scale finite element model. (a) 1st mode; (b) 2nd mode.

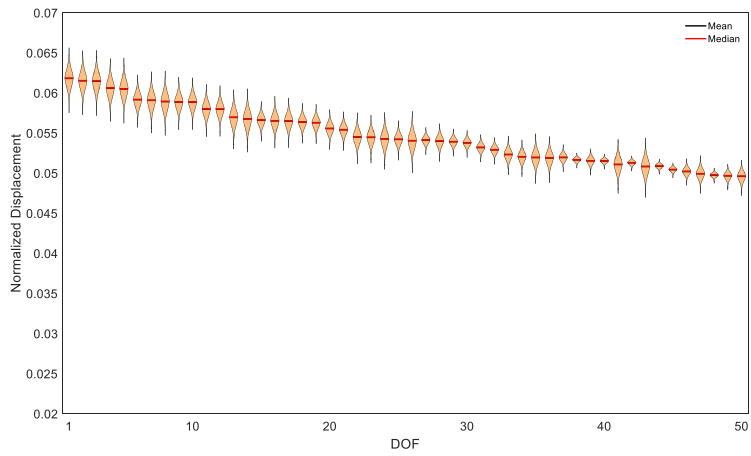

(a)

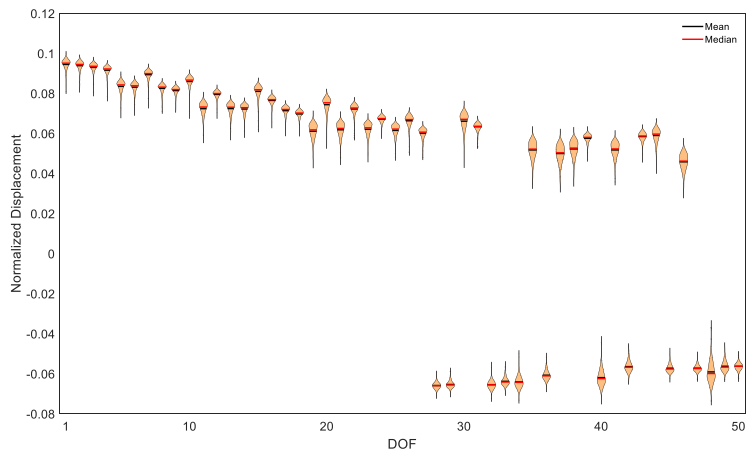

(b)

Figure 4. Distributions of mode shape amplitudes of selected DOFs using Guyan order-reduced model. (a) 1$^{st}$ mode shape; (b) 2$^{nd}$ mode shape.

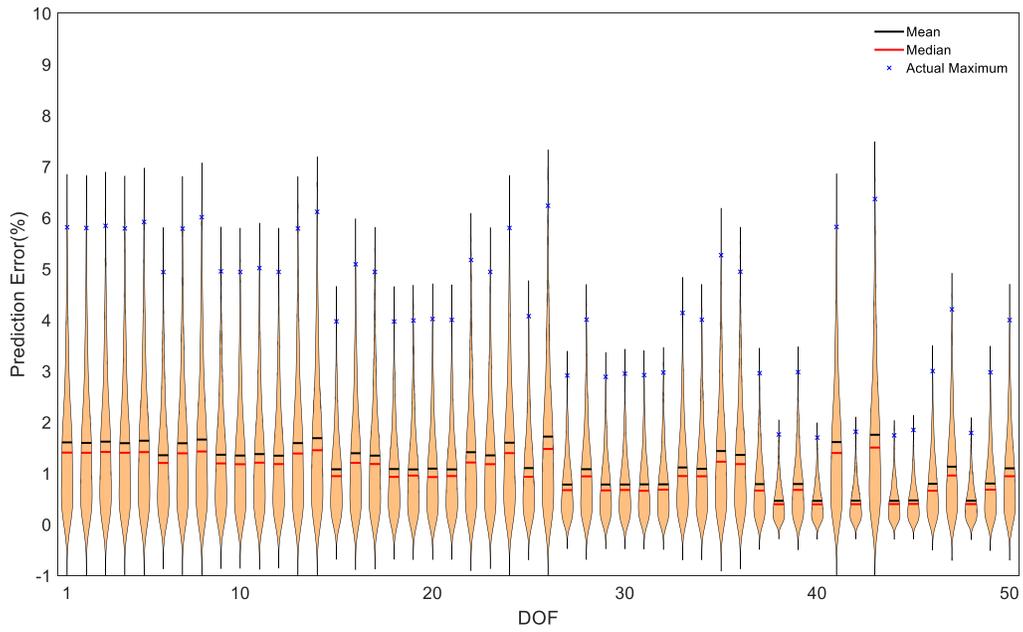

Figure 5. Prediction errors of 1st mode shape amplitudes.

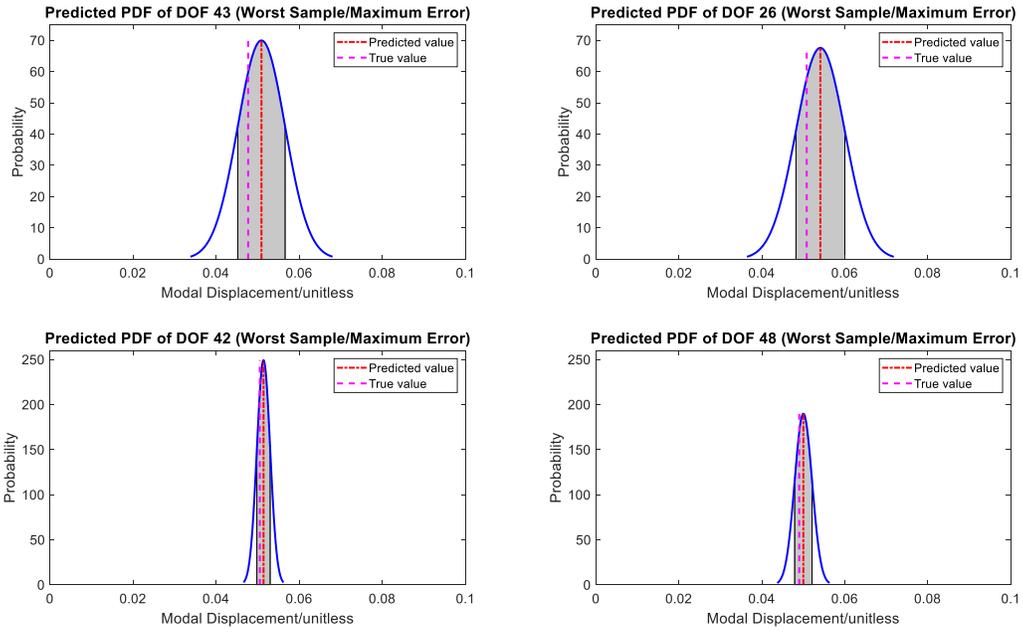

Figure 6. 1st mode shape error analysis based on predicted PDF.

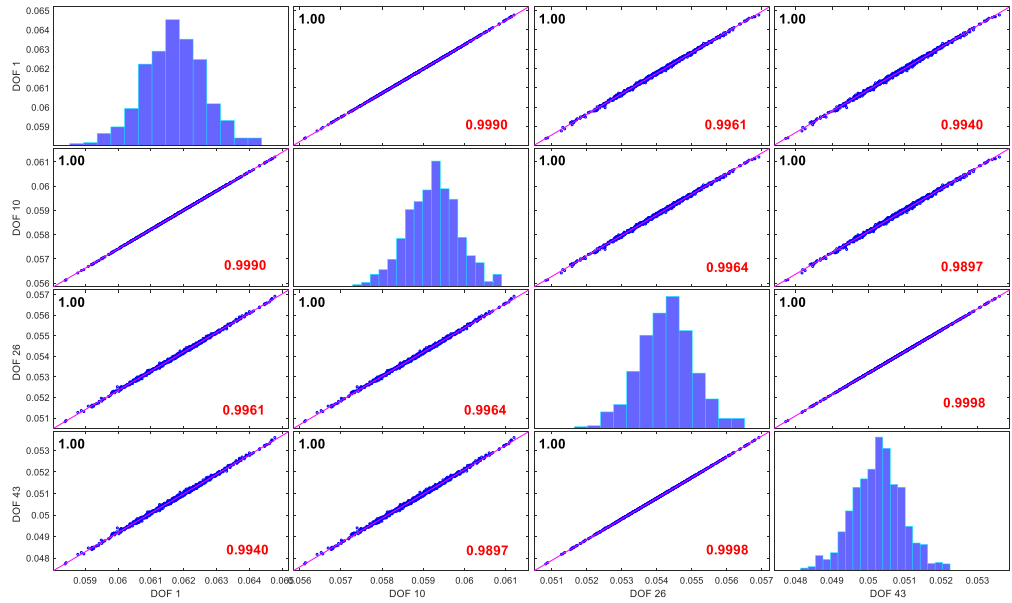

Figure 7. Comparison of output correlation with respect to the original correlation from testing datasets (1$^{st}$ mode shape).

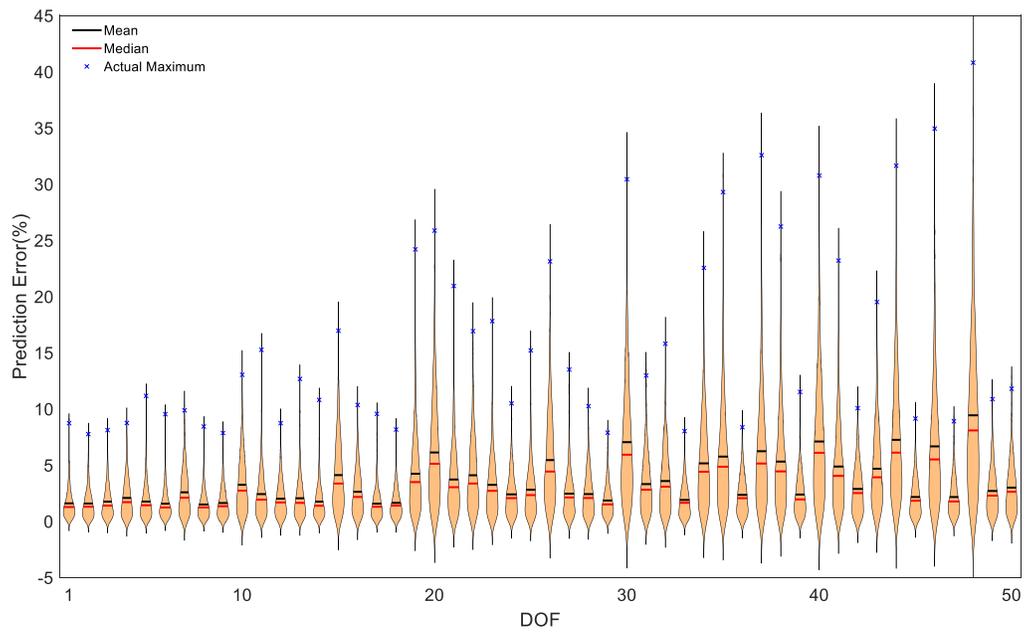

Figure 8. Prediction errors of $2^{nd}$ mode shape amplitudes.

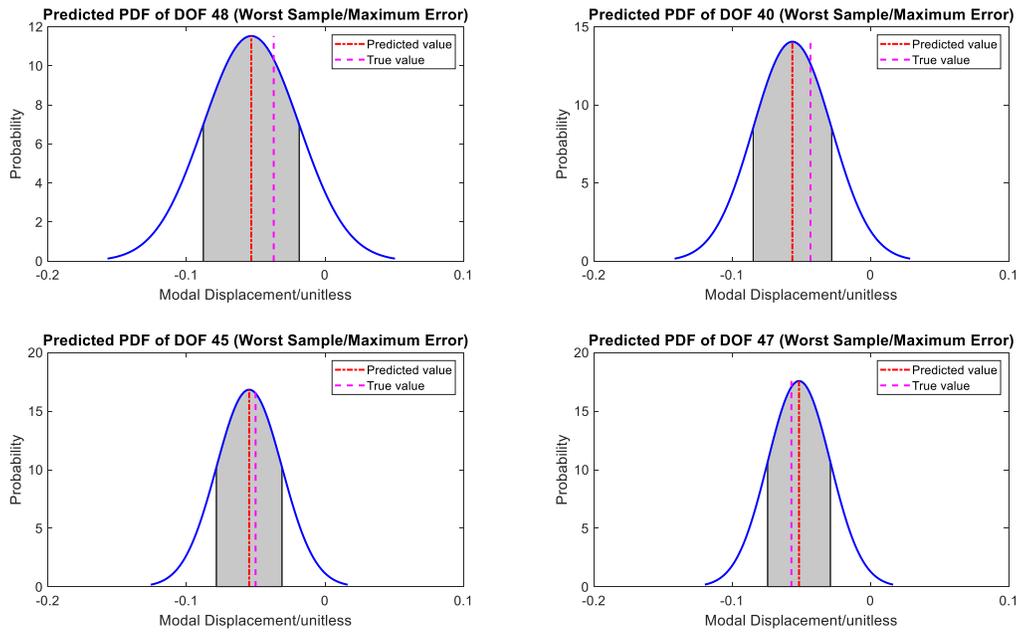

Figure 9. 2nd mode shape error analysis based on predicted PDF.

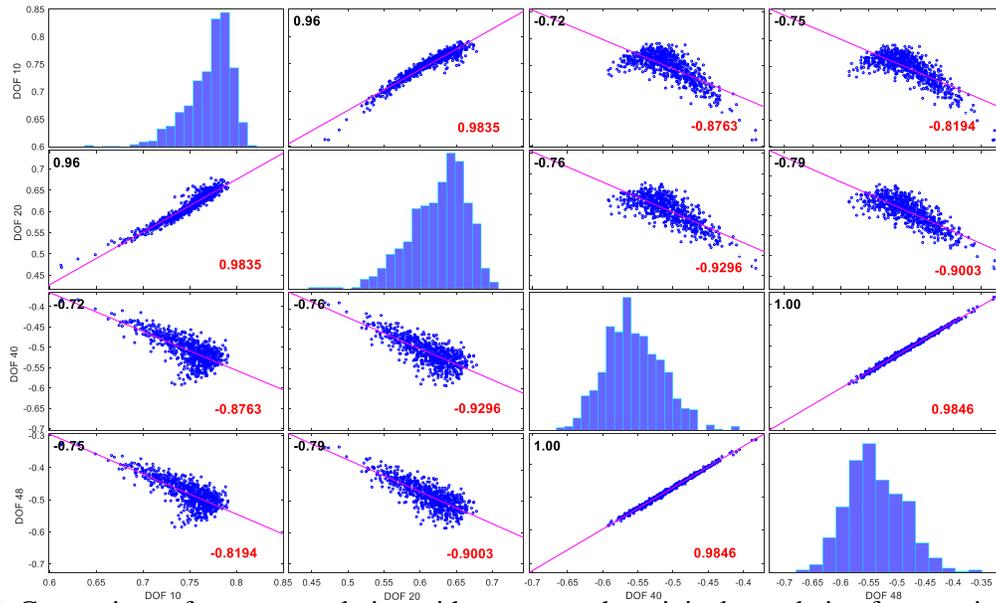
Figure 10. Comparison of output correlation with respect to the original correlation from testing datasets (2$^{nd}$ mode shape).

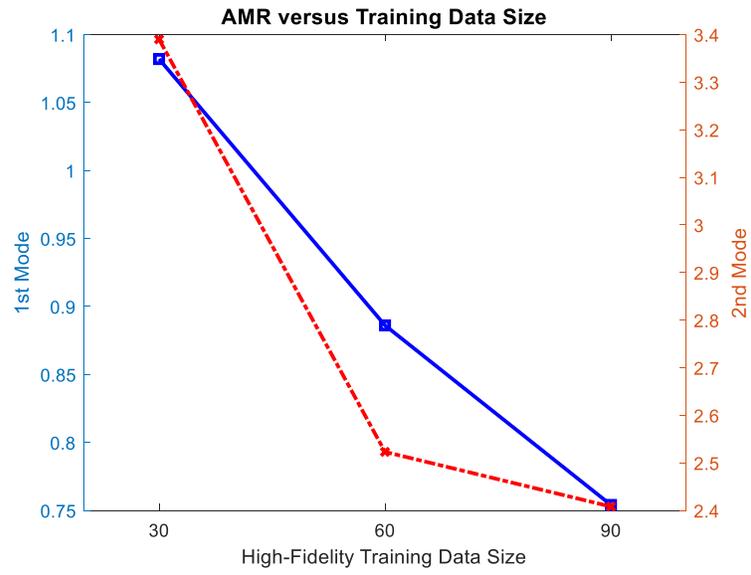

Figure 11. AMR of first two mode shapes versus size of high-fidelity training data.